\documentclass[letterpaper]{article} % DO NOT CHANGE THIS
\usepackage{aaai2026}  % DO NOT CHANGE THIS
\usepackage{times}  % DO NOT CHANGE THIS
\usepackage{helvet}  % DO NOT CHANGE THIS
\usepackage{courier}  % DO NOT CHANGE THIS
\usepackage[hyphens]{url}  % DO NOT CHANGE THIS
\usepackage{graphicx} % DO NOT CHANGE THIS
\usepackage{natbib}  % DO NOT CHANGE THIS AND DO NOT ADD ANY OPTIONS TO IT
\usepackage{caption} % DO NOT CHANGE THIS AND DO NOT ADD ANY OPTIONS TO IT
\usepackage{algorithm}
\usepackage{algorithmic}

\usepackage{newfloat}
\usepackage{listings}
\DeclareCaptionStyle{ruled}{labelfont=normalfont,labelsep=colon,strut=off} % DO NOT CHANGE THIS
\floatstyle{ruled}
\newfloat{listing}{tb}{lst}{}
\floatname{listing}{Listing}
\usepackage{pifont}      % 提供打勾/叉号的符号

\usepackage{graphicx}
\usepackage{booktabs}
\usepackage{array}    % for better column formatting
\usepackage{bm}

\usepackage{makecell}
\usepackage{caption}

\usepackage{xcolor}
\definecolor{customblue}{RGB}{0, 0, 255} 

\usepackage{tabularx}
\usepackage{multirow} 
\usepackage{subcaption}
\usepackage[most]{tcolorbox}

\usepackage{footmisc}
\usepackage{array}
\newcolumntype{M}[1]{>{\centering\arraybackslash}m{#1}}  % 用于垂直居中 + 居中对齐
\newcolumntype{L}[1]{>{\raggedright\arraybackslash}p{#1}}  % 保持右侧定义列不变
\newcolumntype{P}[1]{>{\raggedright\arraybackslash}p{#1}}
\usepackage{placeins}

\title{DisImpact: Quantifying the Physi-Social Impact of Natural Disasters Through Social Media}
\author{
Ruichen Yao\textsuperscript{\rm 1}, 
Tejna Dasari\textsuperscript{\rm 1}, 
Xuanyu Meng\textsuperscript{\rm 1}, 
Elliot Cao\textsuperscript{\rm 1}, 
Zelin Li\textsuperscript{\rm 1}, 
Yifan Liu\textsuperscript{\rm 1}, 
Yaokun Liu\textsuperscript{\rm 1}, 
Lanyu Shang\textsuperscript{\rm 2}, 
Dong Wang\textsuperscript{\rm 1}
}
\affiliations{
    \textsuperscript{\rm 1}University of Illinois Urbana-Champaign, Illinois, USA \\
    \textsuperscript{\rm 2}Loyola Marymount University, California, USA\\
    \textsuperscript{\rm 1}\{ryao8, tejnad2, xuanyum2, elliotc4, zelin3, yifan40, yaokunl2, dwang24\}@illinois.edu; \textsuperscript{\rm 2}lanyu.shang@lmu.edu
}

\usepackage{bibentry}
\begin{document}

\maketitle

\begin{abstract}
Natural disasters not only cause large-scale physical destruction, but also cascading social consequences that are difficult to quantify with traditional surveys and reports. Social media platforms offer an alternative perspective that captures multimodal, real-time, and user-generated content that can be leveraged for disaster impacts. In this paper, we introduce DisImpact, a two-stage framework that systematically quantifies the physi-social impacts of disasters via a Multimodal Large Language Model (MLLM). The social media posts are first classified into ten disaster impact categories that cover both physical and social domains.
We then construct a disaster impact index that integrates the relative prominence of each category with the intensity of public engagement on a weekly basis. This design provides a unified scale for representing disaster impacts across both individual disaster impact categories and the broader physical and social domains. The unified representation enables direct comparison across categories and allows the impacts to be flexibly aggregated to reveal higher-level patterns and overall trends. We validate the impact indices against authoritative ground-truth data, including FEMA Public Assistance data and NASA FIRMS fire detections, observing consistent lead-lag correlations that demonstrate strong validity across both social and physical impact dimensions. We further conduct temporal and spatial analyses, and the results show that physical impacts are often peak during the disasters and localized in regions that are directly affected by disasters, while social impacts often emerge later and spread more broadly across time and space. To the best of our knowledge, this is the first framework to comprehensively quantify disaster impacts across their physical and social dimensions using multimodal data from multiple social media platforms.
\end{abstract}
\section{Introduction}

%---- Cleaned up 1st Section ---% 
The increasing frequency and severity of natural disasters (e.g., hurricanes, wildfires, earthquakes) cause not only large-scale physical destruction but also profound social disruption, posing urgent challenges to human society~\cite{yang_ijcai, yao2025mash,marshall2016mood}. For instance, in September and October 2024, a series of hurricanes struck the United States, together causing more than 250 casualties and more than \$300 billion in economic losses~\cite{ScienceHurricane}. In addition, a dozen of wildfires affected California in January 2025, causing 440 estimated deaths and destroying more than 15,000 houses~\cite{paglino2025excess,Lindsey2025LAFires}.
Beyond physical destruction, disasters also trigger cascading social consequences, ranging from public discourse on resource allocation to heightened perceptions of risk and inequity~\cite{basu2024estimating, li2025quantifying}. 

Traditional documentation of physi-social impacts relies on surveys and government reports. However, these sources often involve significant delays in data collection, capture only a subset of impact dimensions, and are therefore insufficient for capturing the dynamic, real-time evolution of disaster impacts. In contrast, social media posts provide timely, multimodal, and user-driven observations. The posts can reflect both physical consequences, such as damage to infrastructure, resource shortages, and casualties; and social responses, such as distress, blame, and solidarity. Leveraging the real-time social media data enables researchers to move beyond isolated case studies by providing continuous and large-scale observations of how disasters affect communities over time.
% toward systematic measurement of disasters' impact. 
In this paper, we introduce \textit{DisImpact}, a framework that quantifies physi-social impacts of disasters in ten different impact categories, covering both physical (e.g., casualties, damage, resource shortage) and social (e.g., distress, assistance) dimensions, by analyzing multimodal posts from multiple social media platforms. By applying the defined impact categories to posts, our framework captures how different types of impact evolve and interact, offering structured insights that can support disaster response, risk communication, and policy development.

Existing studies have developed multiple ways to quantify the impacts of disasters, but most focus on a single dimension. For example, \citet{miller2025impact} and \citet{mitchell2024mental} measured how exposure to disaster would increase public perceptions of crisis, depression, and suicidal thoughts. \citet{yang2025review} focus on infrastructure disruptions, showing how damage to physical systems undermines community well-being.  \citet{jamal2023understanding} and \citet{basu2024estimating} evaluate preparedness and adaptive capacity through community surveys. 
% Although these approaches provide valuable insights, they assess \orange{one dimension of impact at a time}
% % categories in isolation
% , such as mental health \cite{miller2025impact}, infrastructure loss \cite{yang2025review}, or community resilience \cite{jamal2023understanding}. using domain-specific measures that are not comparable across disaster types, resulting in fragmented assessments that fail to capture how physical and social dimensions interact.
Although these approaches provide valuable insights, they typically examine only one dimension of impact at a time, such as mental-health outcomes measured through clinical symptom scales~\cite{miller2025impact}, infrastructure disruptions quantified through service-interruption metrics~\cite{yang2025review}, or community-resilience indicators derived from social and behavioral data~\cite{jamal2023understanding}. Because these measurements rely on fundamentally different scales, they cannot be directly compared, leading to fragmented assessments that obscure how physical and social impacts relate to each other, as infrastructure damage may amplify psychological distress and existing inequities may determine recovery outcomes.
% This fragmentation limits interconnections between physical and social dimensions, as infrastructure damage may amplify psychological distress and existing inequities often determine recovery outcomes.
% For example, depression rates and infrastructure damage estimates may each be valid within their domains but cannot be combined to assess overall impact. Such siloed measures risk obscuring the interconnected nature of physicosocial impacts: for instance, infrastructure damage can intensify psychological distress, while preexisting inequities shape recovery trajectories. 
In contrast, an integral framework that quantifies both physical and social indicators in a unified scale enables a comprehensive understanding of the impacts of disasters and can provide stronger empirical support for response planning and policy development.

To address these limitations, we propose \textbf{DisImpact}, a two-stage framework that systematically quantifies physi-social impacts 
% via ten disaster impact categories, \orange{covering both physical (e.g.,
% casualties, damage, resource shortage) and social (e.g., distress, as-
% sistance) dimensions,}
using multimodal social media data from multiple platforms. Different from previous studies that lack quantification or examine only a single dimension such as mental health or infrastructure damage, our framework introduces ten impact categories spanning both physical and social dimensions. 
We first leverage a Multimodal Large Language Model (MLLM) to classify a post into defined disaster impact categories based on its multimodal content.
To capture the evolving prominence of each category over time, we then calculate an impact index that combines the proportional prevalence of each category with a measure of public engagement intensity within each time window.
This formulation allows us to capture both the relative prominence of each impact category and the overall intensity of public engagement within each time window, offering a comprehensive and scalable measure of disaster impacts. 
Moreover, as all category-level indices are defined in a unified scale, they function as modular components that can be flexibly combined to construct higher-level composite impact measures, such as physical or social impact dimensions.

To demonstrate the effectiveness of \textbf{DisImpact}, we evaluated it on two large-scale real-world disaster datasets: the \textit{2024 Atlantic Hurricanes} and the \textit{2025 California wildfires}. These events differ in the hazard type, geography, and social context, providing a rigorous testbed for our framework. Through validation studies, DisImpact shows consistent alignment with authoritative ground-truth data: FEMA Assistance data and NASA FIRMS fire detections, indicating that the impact indices can accurately capture both social and physical disaster impacts.
Using DisImpact, we perform both temporal and spatial analysis of the impact indices, revealing that physical impacts peak during the disaster onset, whereas social impacts often rise in the aftermath and diffuse more broadly across time and space.
% Using DisImpact, we perform both temporal and spatial analyses of the impact index, which reveal that physical impacts peak during the disaster onset and are concentrated in directly affected regions, whereas social impacts often rise in the aftermath and diffuse more broadly across time and space. 
% This evaluation demonstrates not only the adaptability of DisImpact across diverse disaster contexts, but also its ability to capture the temporal and spatial divergence between physical and social impacts. 
We envision that the framework can inform a deeper understanding of physi-social impacts, support disaster recovery planning, and enhance real-time disaster management strategies. 
\section{Related Works}
\subsection{Physical and Social Impacts in Disaster}\label{physi-social-impact-review}

% Existing research highlights the impacts of disasters in both the physical realm (casualties, displacement, infrastructure, environment, resources) and social (health, psychological, blame narratives, recovery, socioeconomic) dimensions, both having substantial effects on communities. 
Existing research highlights the impacts of disasters in both the physical realm and social dimensions, both having substantial effects on communities. 

\noindent \textbf{Physical Impacts.}
The most visible effect of disasters is immediate physical destruction and its cascading effects. \textit{Casualties and injuries} rise dramatically from structural collapse, flooding, and entrapment. \citet{parks2021tropical} demonstrat that tropical cyclone exposure increases respiratory illness from mold, smoke, particulates, and contaminated water. These mortality hazards remain elevated for many years after severe flooding, according to a Hurricane study by \citet{keenan2025}. Beyond health risks, disasters force families to evacuate, exposing them to secondary hazards. \citet{jamal2023understanding} highlight that the \textit{evacuation} process itself exposes individuals to dangerous traffic conditions and overcrowded shelters. Severe disasters can cause extensive \textit{damage to infrastructure and utility systems}. \citet{yang2025review} reporte collapsed buildings, flooded roads, and widespread power outages, cutting access to essential lifelines, including food, water, and healthcare. Disasters also cause significant \textit{environmental damage} to the ecosystem. \citet{basu2024estimating} assert that disasters routinely cause unsafe drinking water, dwindling fuel reserves, and depleted medical stockpiles. Breakdowns caused by disasters often cascade into broader \textit{environmental damage}. For example, \citet{juarez2025maui} indicate that wildfires leave behind toxic smoke, devastate agricultural land, and accelerate ecosystem collapse.

\noindent \textbf{Social Impacts.}
Disasters also have a substantial impact on society and local communities. \citet{miller2025impact} show that extreme weather events often result in severe \textit{public health consequences}, including disease outbreaks, mold-related health issues, and disruptions to hospital services. \citet{garfin2022association} demonstrate that repeated exposure to disasters heightens like PTSD, depression, and functional impairment. For example, victims exposed to fires and smoke particles experience increased \textit{emotional and psychological distress}, leading to spikes in mental health visits~\cite{juarez2025maui,jung2025pm25}. In addition, \citet{wertis2023examining} demonstrate that disasters continue to cause \textit{socioeconomic disruptions} such as forced business closures, educational failures, and increased housing insecurity, leading to community destabilization. In addition, disasters also shape social narratives. \citet{abid2024enhancing} highlight that disasters exacerbate \textit{inequities} by fueling political blame, discrimination, or the disproportionate suffering of marginalized populations. \citet{li2025quantifying} show that local community networks, governments, and NGOs collaborate to mobilize \textit{assistance and recovery}, such as providing resources and rebuilding infrastructure, to foster long-term resilience. 

Prior studies examine disaster impacts in narrow classes and rarely quantify them. In this work, we introduce ten physical and social categories and use them to label social media data, producing a holistic and measurable framework.

\subsection{Existing Quantification Methods}

Prior work has quantified disaster impacts using both physical and social measures through distinct, domain-specific methodologies. Physical impacts are commonly assessed through official assessment frameworks such as the Post-Disaster Needs Assessment (PDNA), which aggregates government-led evaluations of damages, losses, and recovery needs using structured surveys, field inspections, and administrative records collected after major events~\cite{GFDRR2017_PDNA_VolA}. Complementing these institutional assessments, \citet{keenan2025} estimate excess mortality following Hurricane Sandy by linking longitudinal Medicare Fee-for-Service beneficiary records with flood exposure at the ZIP Code Tabulation Area level, enabling population-scale inference of disaster-related health effects. \citet{wang2012scalability} introduce a framework to rigorously assess the accuracy of the quantification using estimation theoretical methods.  Social impacts have been examined using a range of survey-based, behavioral, and communication-centered approaches. \citet{garfin2022association} assess post-disaster mental health outcomes among hurricane-exposed populations using standardized post-traumatic stress symptom (PTSS) scales and measures of functional impairment collected through longitudinal surveys. \citet{Mansur2024SocialContracts} analyze public discourse on X during Hurricane Ida using large-scale textual analysis to examine narratives of government responsibility and infrastructure inequality. \citet{wertis2023examining} employ a quasi-experimental design leveraging crisis-text message volumes to measure changes in anxiety, suicidal ideation, and bereavement following Hurricane Ida, enabling causal inference on socioeconomic and mental-health disruption. 
% More recently, social media research has expanded from unimodal text analysis to multimodal classification. \citet{ofli2020mm} introduce the CrisisMMD dataset, in which text and images collected from X are manually annotated into disaster-related categories such as infrastructure damage and rescue efforts. Building on this foundation, \citet{Shetty2025Multimodal} develop multimodal deep-learning models that jointly process text and images to estimate damage severity, demonstrating improved accuracy and granularity over text-only approaches. 
Despite these advances, existing methods remain largely isolated, with most studies focusing on a single impact dimension (e.g., mortality, mental health, or infrastructure damage) and employing distinct measurement pipelines that limit direct comparison across disaster types. Official assessments are systematic but slow and resource-intensive, while social media-based approaches offer faster signals but often rely on a single platform. Our framework addresses these gaps by organizing disaster impacts into ten physical and social categories and introducing a unified index that quantifies both prevalence and intensity across hazards, modalities, and platforms. Leveraging data from Reddit, TikTok, and YouTube, DisImpact expands the representation of public experiences and provides a scalable complement to traditional assessment methods.

\section{Data Collection}

In this study, we focus on two recent disasters, 2024 Pacific Hurricanes and 2025 California Wildfires. For the 2024 Pacific Hurricanes, we focus on the three major hurricanes that hit Florida and North Carolina during the fall of 2024: \textit{Hurricane Francine}, \textit{Hurricane Helene}, and \textit{Hurricane Milton}. According to the National Hurricane Center (NHC), Hurricane Francine was a Category 2 Storm formed on September 9, 2024; Hurricane Helene was a Category 4 Storm formed on September 24, 2024; and Hurricane Miltion was a Category 5 Storm formed on October 5, 2024. The combined damage of these consecutive hurricanes is estimated to be more than \$300 billion~\cite{ScienceHurricane}. For the 2025 California Wildfires, we focus on the wildfires that hit Los Angeles in January 2025. According to NOAA Climate~\cite{Lindsey2025LAFires} and the study of \citet{paglino2025excess}, the 2025 California Wildfires caused 440 estimated deaths and destroyed more than 15,000 houses. To ensure timely and relevant coverage of the disaster-related discussions, we collect posts for hurricanes from September to November 2024, and for wildfires from December 2024 to March 2025. This collection window spans approximately one month before disaster onset and one month after the main impact period, enabling us to capture pre-event signals, peak impacts, and recovery discussions.

We collected posts from three platforms: Reddit, TikTok, and YouTube with a total of 134,053 posts. These three platforms are widely adopted across diverse demographics and communities, enabling the dissemination of information through multiple modalities such as text, images, and videos~\cite{TTandYouTube, TwitterandReddit}. 
The social media data collection adopted the Reddit API, TikTok Research Tools, and YouTube API. 
% The social media data collection adopted the Reddit API\footnote{https://www.reddit.com/dev/api}, TikTok Research Tools\footnote{https://developers.tiktok.com/products/research-api}, and YouTube API\footnote{https://developers.google.com/youtube/v3}. 
We did not include data from X because the platform terminated the free academic API access since 2023 and replaced it with expensive paid tiers, making large-scale data collection financially infeasible for this project~\cite{murtfeldt2024rip, pehlivan2025can, bisiani2025uktwitnewscor}. We utilized different keywords such as \textit{Hurricane Helene}, \textit{Hurricane Milton}, \textit{Los Angeles Wildfire}, and \textit{Palisades Wildfire} to retrieve relevant content for different disasters. 
% In this study, we only consider posts written in English. In addition, we note that social media data may contain users' personal information, raising potential concerns regarding user privacy.
To ensure the ethics of the research and protect users' privacy, we only collected post information (i.e., post ID, post content, post location, and post time) without users' identity information. Tables~\ref{tab:CollectionHurricane} and~\ref{tab:CollectionWildfire} present the number of posts collected from various social media platforms. For text-centric platforms like Reddit, we not only collected the title and description of the post but also collected the attached images and videos. For video-based platforms such as TikTok and YouTube, we collected each post’s video content along with its title and description, while restricting videos to a maximum length of five minutes to reduce the cost and workload for further cleaning and annotation tasks.

\begin{table}[t]
\centering

\setlength{\tabcolsep}{0.5pt}
\footnotesize % 10pt
% {\fontsize{9}{11}\selectfont
\begin{tabular}{cccc}
\toprule
  & \textbf{Reddit}  &  \textbf{TikTok} & \textbf{YouTube} \\
\midrule
Collected Raw Posts    & 12,301   & 67,027           & 2,767           \\
Relevant Posts (\%)   & 9,666 (79\%)                    & 47,921 (71\%)          & 1,707 (62\%)         \\
Avg. \#Words per Rel. Post    &119                    & 21           & 18           \\
\#Images on Rel. Posts    & 1,579                     & --           & --           \\
\#Videos on Rel. Posts & 878                   & 47,921           & 1,707 \\       
\bottomrule
\end{tabular}
\caption{Distribution of Collected Posts on Hurricanes}

\label{tab:CollectionHurricane}
\end{table}

\begin{table}[t]
\centering

\setlength{\tabcolsep}{0.5pt}
\footnotesize % 10pt
% {\fontsize{9}{11}\selectfont
\begin{tabular}{cccc}
\toprule
  & \textbf{Reddit}   & \textbf{TikTok} & \textbf{YouTube} \\
\midrule
Collected Raw Posts    & 9,456   &  40,208          &  2,294          \\
Relevant Posts (\%)    & 6,204 (66\%)                     & 17,567 (44\%)           &  1,339 (58\%)        \\
Avg. \#Words per Rel. Post    & 92                     & 35           & 23           \\
\#Images on Rel. Posts    & 923                      & --           & --           \\
\#Videos on Rel. Posts & 382                      & 17,567           &  1,339 \\       
\bottomrule
\end{tabular}
\caption{Distribution of Collected Posts on Wildfires}

\label{tab:CollectionWildfire}
\end{table}

\begin{table*}[t]
\centering
\footnotesize

\renewcommand{\arraystretch}{1.2}
\begin{tabular}{c|c|L{9cm}}
\toprule[1.5pt]
\textbf{Domain} & \textbf{Category} & \textbf{Definition} \\
\midrule
\multirow{11}{*}{\textbf{Physical Impact}} & \multirow{2}{*}{\makecell{\textbf{Casualties \& Injuries} \\ \textbf{(CINJ)}}} & Posts describing people or animals who are killed, seriously injured, missing, or experiencing immediate medical emergencies. 
\\
\cline{2-3} &
\multirow{2}{*}{\makecell{\textbf{Evacuations \& Displacement} \\ \textbf{(EVAC)}}} & Posts describing people being forced to evacuate, or relocations to shelters caused by unsafe conditions during a disaster.
\\
\cline{2-3} &
\multirow{3}{*}{\makecell{\textbf{Infrastructure \& Utility Damage} \\ \textbf{(INFR)}}} & 
Posts reporting physical damage to infrastructure or utility systems, such as roads, buildings, bridges, or disruptions to essential services such as electricity, water, or communication networks.
\\
\cline{2-3} &
\multirow{3}{*}{\makecell{\textbf{Environmental Damage} \\ \textbf{(ENVD)}}} & Posts describing harm to the natural environment or resources caused by the disaster, such as damage to ecosystems, agriculture, or coastlines, as well as contamination of water, soil, or air.
\\
\cline{2-3} &
\multirow{2}{*}{\makecell{\textbf{Resource Shortages} \\ \textbf{(RSRC)}}} & Posts requesting essential survival resources during a disaster, such as clean water, food, shelter, clothing, or other urgent supplies.
 \\
 \midrule
\multirow{14}{*}{\textbf{Social Impact}} & \multirow{3}{*}{\makecell{\textbf{Public Health} \\ \textbf{(PUBH)}}} & Posts describing public health consequences following a disaster, such as outbreaks of infectious diseases, disruption of chronic illness care, and shortages of medical supplies.
\\
\cline{2-3} &
\multirow{3}{*}{\makecell{\textbf{Emotional \& Psychological Distress} \\ \textbf{(EMOT)}}} & Posts describing psychological distress or emotional suffering experienced by oneself or others as a result of a disaster, such as trauma, anxiety, depression, grief, or cumulative emotional strain.
\\
\cline{2-3} &
\multirow{3}{*}{\makecell{\textbf{Bias Narratives} \\ \textbf{(BIAS)}}} & Posts revealing social bias, discrimination, or unequal impacts of the disaster across different groups, such as political blame, hate speech, or highlighting the disproportionate suffering of vulnerable populations.
\\
\cline{2-3} &
\multirow{4}{*}{\makecell{\textbf{Assistance \& Recovery} \\ \textbf{(ASST)}}} & Posts describing efforts to support recovery and rebuilding after a disaster, such as formal aid from governments or NGOs, informal community-based assistance, long-term infrastructure repair, and expressions of resilience or adaptation by affected communities.
\\
\cline{2-3} &
\multirow{3}{*}{\makecell{\textbf{Socioeconomic Disruption} \\ \textbf{(SECO)}}} & Posts describing economic or social disruptions caused by the disaster, such as financial losses, business interruptions, housing insecurity, educational setbacks, job loss, or the decline of local industries.
\\
\bottomrule[1.5pt]
\end{tabular}
\caption{Definition of Each Disaster Impact Category}

\label{tab:Definition}
\end{table*}

\begin{table}[t]
\footnotesize

    \centering
    \setlength{\tabcolsep}{8pt}
    \begin{tabular}{cccc}
    \toprule
  & \textbf{Reddit}  & \textbf{TikTok} & \textbf{YouTube} \\
\midrule
\textbf{CINJ}    & 332 (3\%)    & 1,207 (3\%)           & 98 (6\%)           \\ 
\textbf{EVAC}    & 368 (3\%)    & 3,212 (6\%)           & 64 (4\%)        \\ 
\textbf{INFR}    & \textbf{1,720 (18\%)}    & \textbf{12,118 (25\%)}           & \textbf{681 (40\%)}           \\ 
\textbf{ENVD}    & 738 (8\%)    & 2,662 (6\%)           & 212 (12\%)           \\ 
\textbf{RSRC}    & 174 (2\%)   & 2,237 (4\%)           & 11 (1\%)           \\ 
\midrule
\textbf{PUBH}    & 155 (2\%)    & 210 (1\%)           & 11 (1\%)           \\ 
\textbf{EMOT}    & 623 (6\%)    & 3,368 (7\%)           & 25 (1\%)        \\ 
\textbf{BIAS}    & 504 (5\%)    & 1,964 (4\%)           & 20 (1\%)           \\ 
\textbf{ASST}    & 866 (9\%)    & \underline{7,048 (15\%)}           & \underline{336 (20\%)}           \\ 
\textbf{SECO}    & \underline{1603 (17\%)}    & 2,414 (5\%)           & 90 (5\%)           \\ 
\midrule

\textbf{OTHER}     & 2,583 (27\%)    & 11,481 (24\%)            & 159 (9\%)           \\ 
    \bottomrule
    \end{tabular}
    % \captionof{table}{Distribution of Physi-Social Impact for Hurricanes}
    \caption{Physi-Social Impact Distribution (Hurricanes)}

\label{tab:SocialDistributionHurricane}
\end{table}

\begin{table}[t]
\footnotesize

    \centering
    \setlength{\tabcolsep}{8pt}
    \begin{tabular}{cccc}
    \toprule
  & \textbf{Reddit}  & \textbf{TikTok} & \textbf{YouTube} \\
\midrule
\textbf{CINJ}    & 144 (3\%)    & 458 (3\%)           & 92 (7\%)           \\ 
\textbf{EVAC}    & 283 (5\%)    & 1,345 (8\%)           & 160 (12\%)        \\ 
\textbf{INFR}    & 541 (9\%)    & 2,745 (16\%)           & \underline{284 (20\%)}           \\ 
\textbf{ENVD}    & \textbf{1312 (21\%)}   & \textbf{5,467 (30\%)}           & \textbf{508 (37\%)}           \\ 
\textbf{RSRC}    & 28 (1\%)    & 80 (1\%)           & 1 (1\%)           \\ 
\midrule
\textbf{PUBH}    & 89 (1\%)    & 336 (2\%)           & 21 (2\%)           \\ 
\textbf{EMOT}    & 204 (3\%)    & 724 (4\%)           & 23 (2\%)        \\ 
\textbf{BIAS}    & 462 (7\%)    & 790 (5\%)           & 15 (1\%)           \\ 
\textbf{ASST}    & \underline{1,137 (18\%)}    & \underline{3,621 (20\%)}           & 162 (12\%)           \\ 
\textbf{SECO}    & 986 (16\%)    & 1,282 (7\%)           & 51 (4\%)           \\ 
\midrule

\textbf{OTHER}     & 1,018 (16\%)    & 719 (4\%)            & 22 (2\%)           \\ 
    \bottomrule
    \end{tabular}
    % \captionof{table}{Distribution of Physi-Social Impact for Wildfires}
    \caption{Physi-Social Impact Distribution (Wildfires)}

\label{tab:SocialDistributionWildfire}
\end{table}

\subsection{Data Cleaning}

We observe that the collected raw social media posts contain irrelevant and off-topic posts. For example, the keyword “hurricane” can appear on posts related to the Miami Hurricanes football team or the Carolina Hurricanes hockey team, while “wildfire” can be used metaphorically to describe rapidly spreading content or emotionally charged discussions. In addition, some posts use disaster-related hashtags solely for promotional purposes, such as adding \#hurricane or \#wildfire to increase exposure and attract more attention to their advertisement. These practices often result in posts with irrelevant content being included in the dataset.

Inspired by \citet{yao2025mash} and \citet{matoshi2025one}, we prompt a Multimodal Large Language Model (MLLM) to label the relevance of each post based on its multimodal content (textual descriptions, images, and videos). We applied separate filtering prompts to each of the hurricane and wildfire datasets, by classifying each post as relevant or irrelevant to pertaining disaster type. Posts that mentioned unrelated events, were removed to ensure that only content related to the disasters was retained. This cleaning process allowed us to reduce the noise from off-topic or promotional posts and focus on the target events, while maintaining scalable volumes of content.
We used Gemini-2.0-flash as the MLLM because of its cost efficiency and advanced capabilities over multimodal inputs~\cite{jegham2025visual, hirosawa2024comparative, team2024gemini}. 
% We used Gemini-2.0-flash as the MLLM because of its cost efficiency\footnote{\label{fn:cost}\blue{Based on historical billing records, the combined API cost for both data cleaning and data annotation was \$350.73 in total.}} and advanced capabilities over multimodal inputs~\cite{jegham2025visual, hirosawa2024comparative, team2023gemini, team2024gemini}. 
The detailed prompt for each disaster can be found in the Appendix Figures~\ref{fig:prompt-clean-hurricane} and~\ref{fig:prompt-clean-wildfire}. In addition, we validated the MLLM-generated annotations through human review on a randomly sampled subset. The results show high agreement between human annotators and between human and MLLM annotations, with detailed results reported in Section \textit{Data Verification}.
% Tables~\ref{tab:CollectionHurricane} and \ref{tab:CollectionWildfire} present the distribution of relevant posts for each social media platform after the cleaning process. 
From Tables~\ref{tab:CollectionHurricane} and \ref{tab:CollectionWildfire}, a total of 59,294 posts were identified as relevant to hurricanes, and 25,110 posts were identified as relevant to wildfires.

\section{Quantify Physi-Social Impact Index}
In this section, we propose a two-stage framework, DisImpact, to quantify the physi-social impacts of disasters.
In the first stage, we prompt an MLLM to annotate each social media post with one of ten predefined impact categories, with five categories representing physical impacts and the other five representing social impacts. The model analyzes the post’s textual and visual content together to assign the most dominant impact category from the predefined set, following the single-label annotation scheme of prior disaster-related social media datasets such as CrisisMMD~\cite{crisismmd} and HumAID~\cite{humaid}. An ``Other” label is included for posts that do not fit any of the ten predefined categories.
% In the second stage, we compute the impact index for each category within each temporal window, based on the distribution of the posts and the level of public engagement. 
In the second stage, we compute category-level impact index for each temporal window by integrating 1) the relative representation of each category within the window and 2) a window-specific weight that reflects the intensity of activity.
% Specifically, we calculate a smoothed proportion of posts per category and multiply it by a window-level weight that reflects the relative intensity of disaster-related discussion during that period. 
The resulting indices reflect both prominence of categories and level of public engagement.

\subsection{Data Annotation}

In the first stage of the DisImpact framework, our goal is to map each social media post to a disaster impact category. We first define ten impact categories informed by prior research, including five categories representing physical impacts and five representing social impacts. The ten disaster impact categories were derived through a multi-stage synthesis process combining prior literature and expert consultation. First, we conducted a structured review of disaster-related studies to extract comprehensive impact concepts in physical and social domains, such as infrastructure damage, environmental degradation, casualties, health outcomes, socioeconomic disruption, and inequality. The detailed summation appears at \textit{Section Physical and Social Impacts in Disaster}. Second, we grouped conceptually similar impacts such as housing damage and infrastructure failure, and political blame and hate speech toward vulnerable groups. Third, experts from interdisciplinary areas were discussed to refine category boundaries and enforce mutual exclusivity at the conceptual level. Impacts were assigned to either the physical or social domain based on whether they primarily reflected material consequences (e.g., casualties, environmental damage, infrastructure loss) or societal and human dimensions (e.g., socioeconomic disruption, inequality, psychological stress).
% These categories are derived by reviewing and synthesizing findings from prior studies on humanitarian needs, community resilience, psychological effects, infrastructure disruptions, and information dynamics in disaster contexts as we summarized in Section~\ref{physi-social-impact-review}~\cite{miller2024impact, yang2025review, jamal2023understanding, basu2024estimating}. 
% We then synthesized these impacts through a series of discussions involving experts from interdisciplinary areas. The experts identified conceptually overlapping impacts, refined ambiguous boundaries, and organized the resulting concepts into ten distinct and mutually exclusive disaster impact categories.
Table~\ref{tab:Definition} presents the list of categories and their definitions. 
% Each category represents a distinct disaster impact observed across prior studies.

To map the content of social media posts to the defined impact category, the most straightforward method is to manually label each post. However, this approach is not only time-consuming and labor-intensive but also financially costly, as our dataset includes over 80,000 relevant posts. To address this challenge, we propose an MLLM-driven approach, supplemented with human verification to assess the reliability of the label. Specifically, the model is prompted to assign the most dominant disaster impact category based on the comprehensive understanding of each post’s multimodal content. While disaster-related posts may in reality reflect multiple co-occurring impacts, the single-label annotation scheme follows prior disaster-related social media datasets such as CrisisMMD~\cite{crisismmd} and HumAID~\cite{humaid} to enable scalable and reproducible annotation at a large-scale dataset. This design provides clear and consistent signals for large-scale annotation, avoiding ambiguity that can arise when multiple overlapping labels are assigned to a single post. To protect user privacy, all tagged usernames, starting with “@”, are replaced with “@user”. The detailed prompt can be found in the Appendix Figure~\ref{fig:prompt-classification}. If the content of the post does not correspond to any of the predefined disaster impact categories, the model is instructed to assign the label as ``Other". Similar to the data cleaning task, we employ Gemini-2.0-Flash as the MLLM due to its strong multimodal reasoning capabilities, rapid inference speed, and low computational cost, which allow efficient processing of large-scale dataset~\cite{jegham2025visual, hirosawa2024comparative, team2024gemini}. To ensure the reliability of the annotations, we randomly sample a subset of posts for manual verification by human annotators from diverse disciplines. As shown in Section \textit{Data Verification}, we observe strong agreement both among human annotators and between human annotators and the MLLM, demonstrating the reliability of the annotations.

Tables \ref{tab:SocialDistributionHurricane} and \ref{tab:SocialDistributionWildfire} present the distribution of disaster impact categories for hurricanes and wildfires, respectively. The largest value in each platform is highlighted in bold and the second-largest value is indicated with underline. We observe both commonalities and divergences across different disasters. For instance, from Tables~\ref{tab:SocialDistributionHurricane} and \ref{tab:SocialDistributionWildfire}, both disasters consistently show high public concern regarding assistance and recovery. This indicates a shared public focus on disaster support and community rebuilding, reflecting a fundamental dimension of societal response that transcends specific disasters. In addition to the commonalities, we also identify differences between disasters. As shown in Table \ref{tab:SocialDistributionHurricane}, discussions during hurricanes focus more on infrastructure damage, whereas Table \ref{tab:SocialDistributionWildfire} reveals that environmental damage is more frequently mentioned during wildfires. This divergence is likely attributed to the pervasive dispersion of wildfire smoke, which can impact regions far beyond the immediate burn zones, thereby elevating broader environmental concerns compared to the typically localized effects of hurricanes. Moreover, platform-level differences also emerge. For example, discussions of socioeconomic disruption appear more prominent on Reddit compared to other platforms. 
% In addition, posts on X contain a higher proportion of bias-related narratives, indicating potential differences in user behaviors across platforms.

% Figure \ref{fig:physisocial_count_trends} shows the 

\subsection{Data Verification} \label{DataVerification}
To verify the reliability of the annotation, we invite three human annotators across different disciplines to manually verify the results, similar to the work by \citet{manakul2023selfcheckgpt, liu2025reasoning}. Specifically, we conducted human verification for two tasks: data cleaning and disaster impact classification.
For the data cleaning task, we randomly sampled a subset of posts from the raw collected dataset for each annotator. For the disaster impact classification task, we separately sampled posts from the cleaned relevant dataset.
In both tasks, we sampled 150 posts for each disaster–platform pair, including 50 overlapping posts for measuring inter-annotator agreement. This sampling scheme results in 900 posts per annotator for each task, corresponding to all combinations of two disasters and three platforms. To verify the agreement of annotations between human annotators, we adopted two metrics, Consistency and Fleiss’ Kappa $\kappa$ Score, for evaluation. Consistency calculates the proportion of samples that are completely consistent among all annotators, while Fleiss’ Kappa Score calculates the overall consistency after random consistency correction between annotators~\cite{fleiss1971measuring}.
Let $A$, $B$, and $C$ be the labels assigned by each annotator, 
$i$ represents the index of the data sample, and $N$ is the total number of data samples. 
The Consistency is defined as: 
% $\text{Consistency} = \frac{\sum_{i=1}^{N} \mathbf(A_i = B_i = C_i)}{N}$.
{\small
\begin{equation}
\text{Consistency} = \frac{\sum_{i=1}^{N} \mathbf(A_i = B_i = C_i)}{N}
\end{equation}}
Let $\bar{P}$ denote actual agreement (i.e., the average proportion of actual agreement between annotators) and $\bar{P_e}$ denote the expected agreement (i.e., the expected agreement if annotators randomly choose categories). The Fleiss' Kappa Score is defined as: 
{\small
\begin{equation}
    \text{Fleiss'} \ \kappa = \frac{\bar{P} - \bar{P}_e}{1 - \bar{P}_e}
\end{equation}}
Subsequently, we compute Consistency and Cohen’s Kappa $\kappa$ Score to evaluate the agreement between human consensus and MLLM's annotations. Different from Fleiss’ Kappa, Cohen’s Kappa accounts for chance agreement between two raters 
% by correcting for the proportion of agreement 
and is therefore more appropriate for assessing pairwise annotator reliability~\cite{cohen1960coefficient}.
Let $P_o$ denote the observed agreement between two raters, and $P_e$ denote the expected agreement by chance. The Cohen’s Kappa score is defined as: 
% $\text{Cohen’s} \ \kappa = \frac{P_o - P_e}{1 - P_e}$.
{\small
\begin{equation}
\text{Cohen’s} \ \kappa = \frac{P_o - P_e}{1 - P_e}
\end{equation}}

% Subsequently, we evaluate the agreement between MLLM-generated labels and human annotations using Consistency and Cohen’s Kappa $\kappa$ Score, following the same metrics applied in assessing inter-annotator agreement.
Table~\ref{tab:humananno} presents the inter-annotator agreement, as well as the agreement between human consensus and the MLLM generated outputs. For both the data-cleaning task and the disaster-impact classification task, the results demonstrate high agreement among human annotators as well as between humans and the MLLM, reflecting the reliability of MLLM.

\begin{table}[t]
\centering
\setlength{\tabcolsep}{1pt}
\footnotesize % 10pt
\begin{tabular}{ccc||cc}
\toprule
  & \makecell{\textbf{Human-Human} \\ \textbf{Consistency}} & \textbf{Fleiss’ $\kappa$} & \makecell{\textbf{Human-MLLM} \\ \textbf{Consistency}}& \textbf{Cohen’s $\kappa$} \\
\midrule
\textbf{H-Clean} & 0.9067 & 0.6794  & 0.9533 & 0.7828 \\
\textbf{W-Clean} & 0.7533 & 0.6450  & 0.9100 & 0.8112 \\
\midrule
\textbf{H-Anno} & 0.7133 & 0.7796  & 0.9056 & 0.8895 \\
\textbf{W-Anno} & 0.7267 & 0.7816 &  0.8800 & 0.8621 \\
\bottomrule
\end{tabular}
\caption{Annotation Verification Results. H stands for Hurricane, and W stands for Wildfire.}
\label{tab:humananno}
\end{table}

\begin{table*}[t]
\centering
\footnotesize
\begin{tabular}{cccccccc}
\toprule
\textbf{Indicator–Disaster} & \textbf{-3 week} & \textbf{-2 week} & \textbf{-1 week} & \textbf{0 week} & \textbf{+1 week} & \textbf{+2 week} & \textbf{+3 week} \\
\midrule
FEMA--Hurricane (Impacts in Social Domain)   & 0.440 & 0.214 & -0.008 & -0.309 & -0.528 & -0.552 & -0.466 \\
FIRMS--Wildfire (Impacts in Physical Domain) & 0.105 & -0.228 & -0.289 & 0.424 & -0.546 & 0.533 & -0.333 \\
\bottomrule
\end{tabular}
\caption{Lead-Lag Spearman Correlations with Authorized Ground-truth Data}
\label{tab:valid}
\end{table*}

\begin{figure*}[ht]
    \centering
    \begin{minipage}{1\textwidth} % 设置整个区域的宽度
        \centering
        \includegraphics[width=0.9\textwidth]{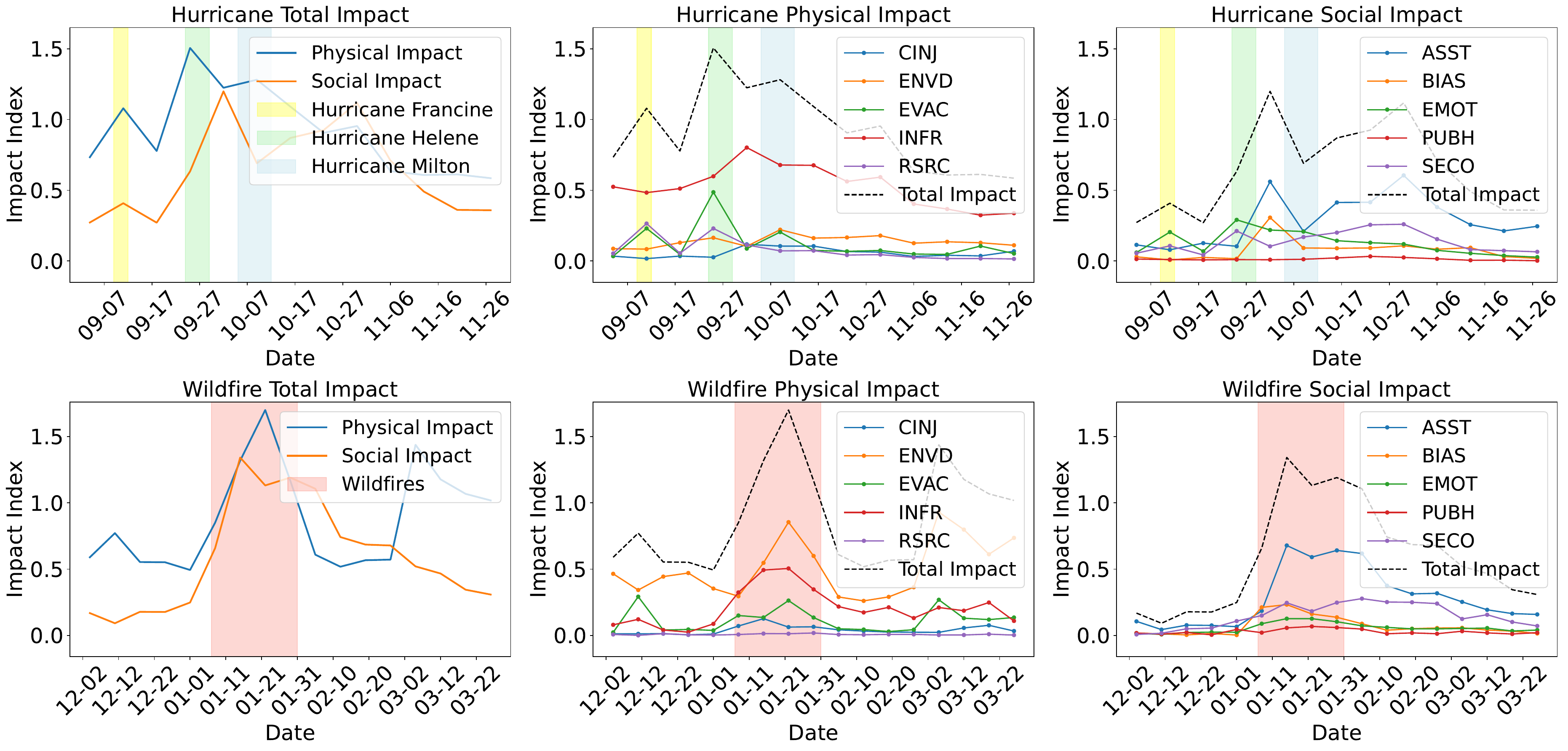} % 图片宽度限制为 minipage 宽度
        \caption{Physical and Social Impact Over Time} % 设置标题
        \label{fig:physisocial_impact_trends} % 设置图片引用标签
    \end{minipage}
\end{figure*}

\subsection{Impact Index Calculation}

In the second stage of the DisImpact framework, we design a method to calculate the intensity of each disaster impact category within fixed-length time windows, based on the category labels assigned to individual posts during the first stage.
Following the design in~\citet{shang2025side}, we segment the timeline into weekly intervals (7 days), allowing for consistent temporal aggregation and comparison across disaster phases.
% We first calculate the impact index for each disaster impact category over time. Specifically, w
We first calculate the smoothed proportion $P$ (in the interval (0, 1)) of posts assigned to each category $c$ within a given time window $t$ to capture the relative prominence of different impact categories:
{\small
\begin{equation}
    P_t(c) = \frac{n_t(c) + \alpha}{N_t + \alpha \cdot C}
\end{equation}
}

\noindent where $n_t(c)$ is the number of posts classified into the category $c$ during the time window $t$, $C$ is the total number of categories, and $N_t = \sum_{c=1}^{C} n_t(c)$ is the total number of posts from all categories during time window $t$. The number of posts within a given window can fluctuate greatly depending on the disaster’s severity. For example, categories with relatively small absolute counts may be overshadowed during periods of intense public discourse, whereas rare categories may exert a large influence during periods of limited activity due to inflated relative proportions, which distorts the overall distribution.
To address this challenge, we adopt additive smoothing instead of raw proportions. The additive smoothing ensures that every category maintains a nonzero contribution across time windows, thereby stabilizing the distributions and mitigating the impact of fluctuations. Following the Jeffreys prior convention~\cite{jeffreys1998theory} and the study of \citet{chen1999empirical}, we set the smoothing parameter $\alpha = 0.5$ to correct the instability due to small counts while avoiding over-smoothing.

Subsequently, we compute the intensity weight $w_t$ for each time window $t$ to modulate the impact index based on the level of discussion activity. The goal of introducing intensity weight is to amplify the influence of high-volume periods while attenuating the impact of windows with sparse discussions, which are more prone to unstable variations. The design of weight is grounded in the intuition that periods of intense public discourse tend to reflect stronger or more widespread societal impacts, whereas low-activity periods, even if dominated by a specific category, may not signify a meaningful or large-scale effect. 
During the active phase of a disaster, public attention typically intensifies and social media platforms exhibit substantial discussions across multiple impact-related categories. In such high-activity windows, even categories with small relative proportions can correspond to large absolute volume of posts, and should therefore contribute more strongly to the impact index. Conversely, in later stages when overall discussion volumes decline, a category may dominate in a time window with a small number of remaining posts. Such dominance only reflects isolated or residual conversations rather than sustained or widespread impact. By incorporating an intensity weight that reflects the overall discussion volume in each time window, the index avoids overemphasizing noisy low-volume windows or underestimating categories solely due to small relative proportions, thereby reducing the risk of obscuring key impact patterns and ensuring consistency in the resulting index values.
% The goal of introducing intensity weight is to amplify the influence of high-volume periods (e.g., peaks in public attention) while attenuating the impact of windows with sparse discussions, which are more prone to spurious fluctuations. The design of weight is grounded in the intuition that periods of intense public discourse tend to reflect stronger or more widespread societal impacts, whereas low-activity periods, even if dominated by a specific category, may not signify a meaningful or large-scale effect. For instance, during the active phase of the disaster, public attention typically intensifies and social media platforms exhibit substantial discussion volumes across many impact-related topics. These high-activity periods are more likely to exert strong influences and widespread effects and should therefore contribute more to the impact index. In contrast, the overall discussion volumes decline several weeks after the disaster. Even if a particular category dominates the small number of remaining posts, such late-stage proportions often reflect limited or isolated discussions rather than a sustained or large-scale impact. The intensity weight helps prevent these low-volume windows from generating spurious spikes in the index.
The intensity weight is defined as:

{\small
\begin{equation}
     w_t = \arctan\left(\frac{N_t - N_{mean}}{\text{IQR}}\right) + \frac{\pi}{2}
\end{equation}}

\noindent Specifically, we first measure the deviation between the total number of posts $N_t$ in window $t$  and the overall mean posting volume $N_{\text{mean}} = \frac{1}{T} \sum_{t=1}^{T} N_t$, and normalize the difference using the interquartile range (IQR) of post counts across all time windows.
We then apply the arctangent function as a soft activation, which smoothly distinguishes between low- and high-activity periods while preventing sharp spikes caused by extreme outliers.
Finally, we add an offset of $\frac{\pi}{2}$ to ensure that all weights remain positive, thereby keeping the resulting impact index values within the interval (0, $\pi$). In practice, $w_t$ approaches 0 when the posting volume is far below the mean ($N_t \ll N_{\text{mean}}$), approaches $\pi$ when posting volumne is far above the mean ($N_t \gg N_{\text{mean}}$), and takes the mid-range value $\frac{\pi}{2}$ when posting volumne approximately equals the mean ($N_t \approx N_{\text{mean}}$).
The impact index for category $c$ in time window $t$ is defined as the product of the smoothed proportion $P_t(c)$ and the intensity weight $w_t$:
{\small
\begin{equation}
    I_t(c) = P_t(c) \cdot w_t
\end{equation}}

\noindent This formulation ensures that the impact index captures both the relative prominence of category $c$ within the window and the overall level of public attention during that period, allowing us to jointly account for content-specific importance and temporal discussion intensity. As a result, the final index reflects both the absolute level of activity within a window and the relative prominence of each impact category. Moreover, because all category-level indices are expressed on a unified scale, they can be flexibly combined as modular components to construct higher-level physical and social impact measures in a modular manner.

%%%%%%%%%%%%%%%%%%%%%%%%%%%%%%%%%%%%%%%%%%%%%%%%%%%%%%%%%%%%%%%%%%%%%%%%%%%%%%%%%%%%%

\begin{figure*}[ht]
    \centering
    \begin{minipage}{1\textwidth} % 设置整个区域的宽度
        \centering
        \includegraphics[width=0.9\textwidth]{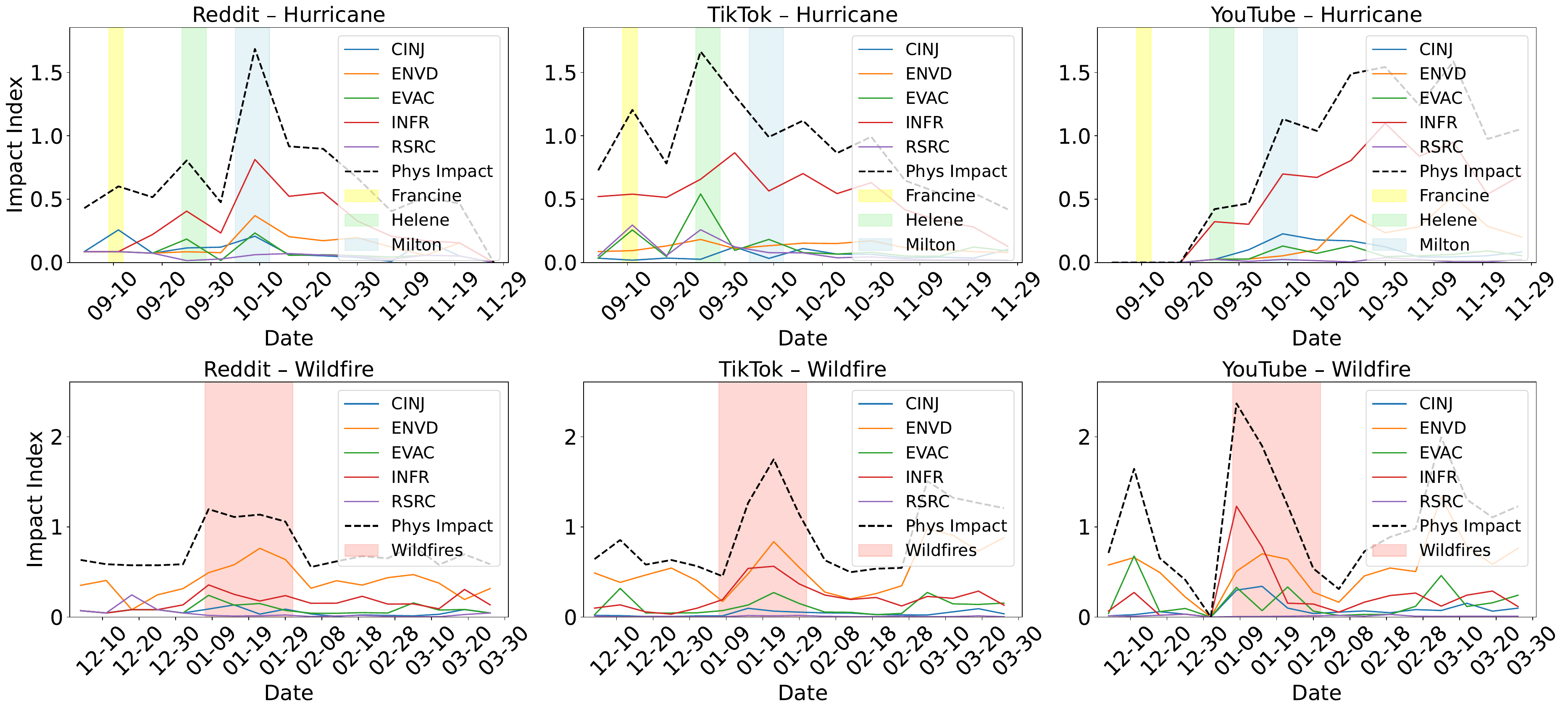} % 图片宽度限制为 minipage 宽度
        \caption{Physical Impact Over Time for each Platform} % 设置标题
        \label{fig:physical_impact_trends} % 设置图片引用标签
    \end{minipage}
\end{figure*}

\begin{figure*}[ht]
    \centering
    \begin{minipage}{1\textwidth} % 设置整个区域的宽度
        \centering
        \includegraphics[width=0.88\textwidth]{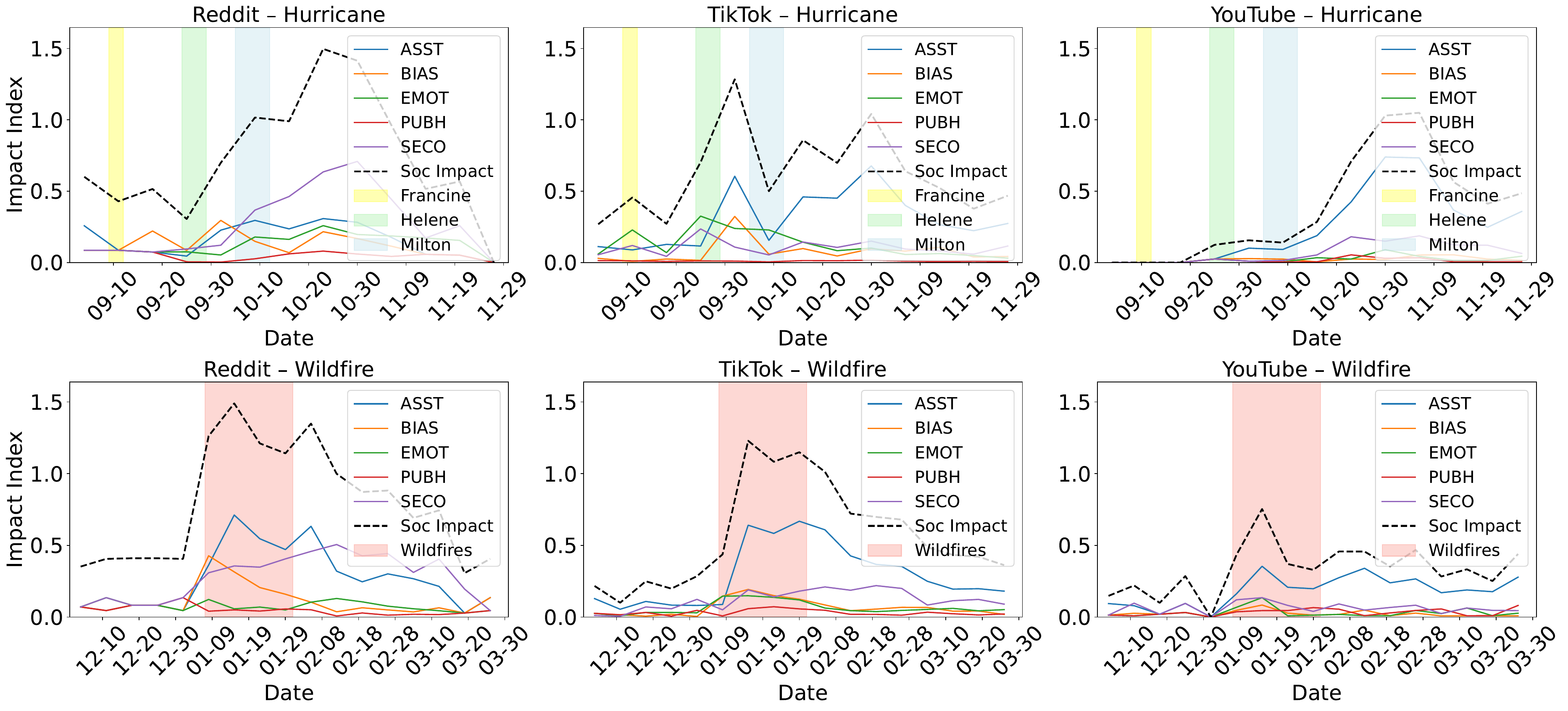} % 图片宽度限制为 minipage 宽度
        \caption{Social Impact Over Time for each Platform} % 设置标题
        \label{fig:social_impact_trends} % 设置图片引用标签
    \end{minipage}
\end{figure*}

\section{Impact Index Analyses}

% To validate the reliability of the impact index and better understand the patterns of disaster-related discourse, we perform impact index validation and temporal analysis
% % and spatial analyses 
% on the computed impact indices for both hurricane and wildfire datasets. 
% We implement impact index validation and temporal analysis in this section.
% Due to page limitations, we perform spatial analysis in Appendix~\ref{spatial-analysis}.

\subsection{Impact Index Validation}

% To validate DisImpact, we compare our indices against authoritative ground-truth sources selected for construct validity: physical indices against physical damage metrics, and social indices against community-level recovery indicators. 
% % NOAA Storm Damage Estimates provide official post-storm assessments of U.S. hurricane damage based on engineering surveys, insurance data, and local government reports; these dollar-denominated estimates serve as the standard reference for physical impacts. 
% FEMA Public Assistance Allocations distribute federal recovery funds after declared disasters and offer the closest available proxy for community social needs. NASA FIRMS Fire Detections supply satellite-based measurements of active fire radiative power from MODIS/VIIRS instruments, providing objective indicators of wildfire intensity. 
% % The EPA’s Air Quality Index (AQI) measures pollutant concentrations and, during wildfire events, captures periods of acute community stress and disruption; AQI serves as the primary ground-truth source for validating wildfire-related Social Impact indices.
% All sources are aggregated to weekly intervals to align with our social-media collection window, with NOAA and FEMA data obtained from official government repositories and FIRMS and AQI retrieved via public APIs.

To validate DisImpact, we compare our indices against authoritative ground-truth sources selected for construct validity. 
% DisImpact is validated as a construct-aligned signal rather than as a predictive estimator. Therefore, we do not expect monotonic alignment across all temporal lags, but interpret systematic lead–lag structure as reflecting institutional reporting delays and physical immediacy in external indicators.
Specifically, to reflect community-level social needs, we select FEMA Public Assistance Allocations data, which capture federal recovery funding distributions following disaster events. To represent objective indicators of wildfire intensity, we select NASA FIRMS fire detection data, which provide satellite-based measurements of active fire radiative power. Both external signals are aggregated to weekly intervals to align with the temporal resolution of our impact indices. In particular, we compare social impact indices derived from hurricane-related social media data with FEMA Public Assistance allocations, and physical impact indices derived from wildfire-related data with FIRMS fire-radiative-power measurements.

We use the Spearman rank correlation coefficient ($\rho$) to quantify monotonic association between the calculated impact indices and ground-truth measurements. Spearman $\rho$ is suitable because it captures monotonic trends without assuming linearity, tolerates scale differences, and is robust to outliers and non-Gaussian distributions~\cite{schober2018correlation}. To account for timing differences between real-time social media activity and delayed official reporting, we compute correlations at temporal offsets of $\pm 3$ weeks: negative lags ($\ell < 0$) indicate social media leads, zero lag ($\ell = 0$) indicates concurrent movement, and positive lags ($\ell > 0$) indicate ground truth leads. For each temporal lag $\ell$, we correlate the impact index in week $t$ with the ground-truth measurement in week $t+\ell$, allowing us to evaluate how social-media signals align with, lead, or lag behind official indicators. The lead-lag correlation results are available at Table~\ref{tab:valid}. Following conventions in disaster informatics, correlations in the $0.3$-$0.5$ range are typically viewed as meaningful due to the inherent cross-domain noise between social and physical indicators~\cite{kryvasheyeu2016rapid}.

From Table~\ref{tab:valid}, the social impact indices demonstrate construct validity when compared with FEMA Public Assistance data, with the strongest association observed at a 3-week lead. This lead–lag trend indicates that the social impact indices systematically precede FEMA disbursements by approximately three weeks, reflecting its ability to capture emerging community needs earlier than official financial processes. The negative contemporaneous correlation further aligns with the documented delays inherent in FEMA’s recovery-phase funding processes. In addition, the physical impact indices show moderate construct validity when evaluated against FIRMS fire-radiative-power measurements. The strongest contemporaneous association occurs at concurring, indicating that physical impact indices can capture real-time wildfire intensity as measured by FIRMS. Together, the hurricane and wildfire validation results provide convergent evidence that the impact indices can produce timely, meaningful, and externally grounded measurements of both social and physical disaster impacts.
% In essence, we assess whether social media functions as a reliable “thermometer” for real-world disasters: if weekly DisImpact scores rise and fall in alignment with ground-truth data—whether in the same week or at a lead/lag offset—we consider the index validated. 

% \begin{figure}
%     \centering
%     \includegraphics[width=\linewidth]{20539851-ca45-408e-a444-e9f84ebd06e6.jpeg}
%     \caption{Enter Caption}
%     \label{fig:placeholder}
% \end{figure}

% Hurricane Social Impact   FEMA: Social impact indices show moderate validation against FEMA allocations, with the strongest correlation at lag −3 weeks. This suggests social media signals lead FEMA disbursements by about three weeks. The negative concurrent correlation (ρ=−0.309) reflects FEMA’s delayed recovery-phase funding, and the consistent lead pattern indicates DisImpact captures emerging social needs earlier than official metrics.

% Wildfire Physical Impact   FIRMS: Physical impact indices moderately validate against FIRMS fire radiative power, with a peak correlation at lag +2 weeks and solid concurrent alignment (ρ=0.424). The positive lag reflects slower social media documentation relative to peak fire activity, while the concurrent correlation shows DisImpact reliably tracks real-time wildfire conditions.

\subsection{Temporal Analysis}

We aggregate posts from all social media platforms to investigate the comprehensive impact of each disaster.
% Figure~\ref{fig:physisocial_impact_trends} illustrates how each disaster impact category evolves over time for both hurricanes and wildfires.
% We further aggregate the impact indices of categories within the physical and social domains to compute the overall physical impact and social impact, respectively.
Figure~\ref{fig:physisocial_impact_trends} reports the aggregated physical and social impact indices for hurricanes and wildfires. We subsequently decompose these aggregated trends by illustrating the temporal dynamics of each disaster impact category.
We observe that both the physical and social impact indices show increases during the core disaster periods.
This temporal alignment between peak impact values and known disaster timelines provides support for the validity of the disImpact framework.
It indicates that our method is capable of capturing time-sensitive signals that reflect shifts in public attention, concern, and discourse patterns during disasters.

In addition, we observe that the physical impact index decreases rapidly after the disaster, while the social impact index exceeds the physical impact index and remains at a relatively high level after hurricanes. This transition suggests that after the immediate physical consequences of the disasters, public attention shifts more toward social dimensions such as recovery, inequality, and socioeconomic disruption. A similar pattern emerges in the wildfire case, where the social impact index temporarily exceeds the physical impact index after the disaster. However, in contrast to the hurricane case where the physical impact steadily declines after the disaster, the physical impact index in the wildfire case shows a resurgence after an initial decrease. A closer examination of category-level trends reveals that this rebound is primarily driven by a noticeable increase in the environmental damage index. This resurgence in the physical impact index is likely driven by increased public concern over the widespread dispersion of wildfire smoke, which brings renewed attention to environmental damage and broader physical consequences, even after the fire itself has subsided.

Subsequently, we compute the disaster impact index for each category separately for each platform, allowing us to analyze both the convergent and divergent patterns of category-level impact across platforms. Figures \ref{fig:physical_impact_trends} and \ref{fig:social_impact_trends} present the impact index for categories in the physical and social domains, respectively. For categories within the physical domain, we observe a high degree of consistency across platforms in terms of impact index patterns.
During hurricanes, infrastructure damage stands out as one of the most prominent physical impact categories, indicating a shared emphasis on disruptions such as road blockages, power outages, and building damage.
Similarly, in the wildfire dataset, environmental damage emerges as the most dominant physical impact category, reflecting a common concern with ecological consequences such as smoke and habitat loss.

However, for social-domain categories, we observe more pronounced differences across platforms.
For instance, the socioeconomic impact index on Reddit is substantially higher than that of other platforms, suggesting that Reddit users tend to engage more with topics related to economic hardship, job loss, and resource access.
% In contrast, bias-related narratives exhibit a significantly elevated impact index on X, indicating a heightened presence of biased charged discourse.
These findings highlight that different platforms may emphasize different facets of social impact, shaped by their respective user bases, content norms, and communicative affordances.
This observation underscores the critical importance of incorporating data from multiple social media platforms, as relying on a single platform may result in an incomplete or biased representation of public discourse.

\subsection{Spatial Analysis} \label{spatial-analysis}

\begin{table*}[t!]
\centering
\footnotesize
\begin{tabular}{c cc|cc|cc}
\toprule
 & \multicolumn{2}{c|}{\textbf{Reddit}} 
 & \multicolumn{2}{c|}{\textbf{TikTok}} 
 & \multicolumn{2}{c}{\textbf{YouTube}} \\
\cmidrule(lr){2-3} \cmidrule(lr){4-5} \cmidrule(lr){6-7}
 & Hurricane & Wildfire 
 & Hurricane & Wildfire 
 & Hurricane & Wildfire \\
\midrule
% Relevant Posts 
%  & 9,844 & 6,204
%  &  47,921 & 17,567
%  & 1,707  &  1,339 \\
%  \midrule
Posts with Location from Metadata
& 0 (0\%) & 0 (0\%)
& 13,903 (38\%) & 4,061 (24\%)
& 243 (14\%) & 106 (8\%)\\
Posts with Location from Textual Content 
& 1,374 (19\%) & 2,014 (39\%)
& 7,707 (21\%) & 6,945 (41\%)
& 524 (31\%) & 623 (47\%)\\
\midrule
Total Number of Posts with Location
& 1,374 (19\%) & 2,014 (39\%)
& 21,610 (59\%) & 11,006 (65\%)
& 767 (45\%) & 729 (55\%)\\
\bottomrule
\end{tabular}
\caption{Distribution of Location Information}
\label{tab:location}
\end{table*}

To examine the spatial dimension of the disaster impact index, we employ two approaches to infer the geographic location of each post.
The first way is to directly utilize location metadata if the post was explicitly provided.
The second way applies to posts lacking metadata-based geolocation, following prior works~\cite{yao2025mash,vracevic2025altgeosocial} in social media analysis. The distribution of posts labeled by each location method across is reported in Table~\ref{tab:location}.
% From Table~\ref{tab:location}, we observed that relying on location data from a single method would lead to sparse spatial coverage. For example, Reddit posts do not provide metadata-based geolocation information, and YouTube posts only contain a limited fraction of metadata-based locations. As a result, computing impact indices and conducting spatial analysis based on a single geolocation method would be constrained by data sparsity and could introduce bias. Therefore, we integrate the posts with geolocation from both methods for the spatial analysis.} 
Posts for which neither method yields a valid geographic reference are excluded from the spatial analysis.

\begin{figure}[t]
    \centering
    \begin{minipage}{0.5\textwidth} % 设置整个区域的宽度
        \centering
        \includegraphics[width=0.82\textwidth]{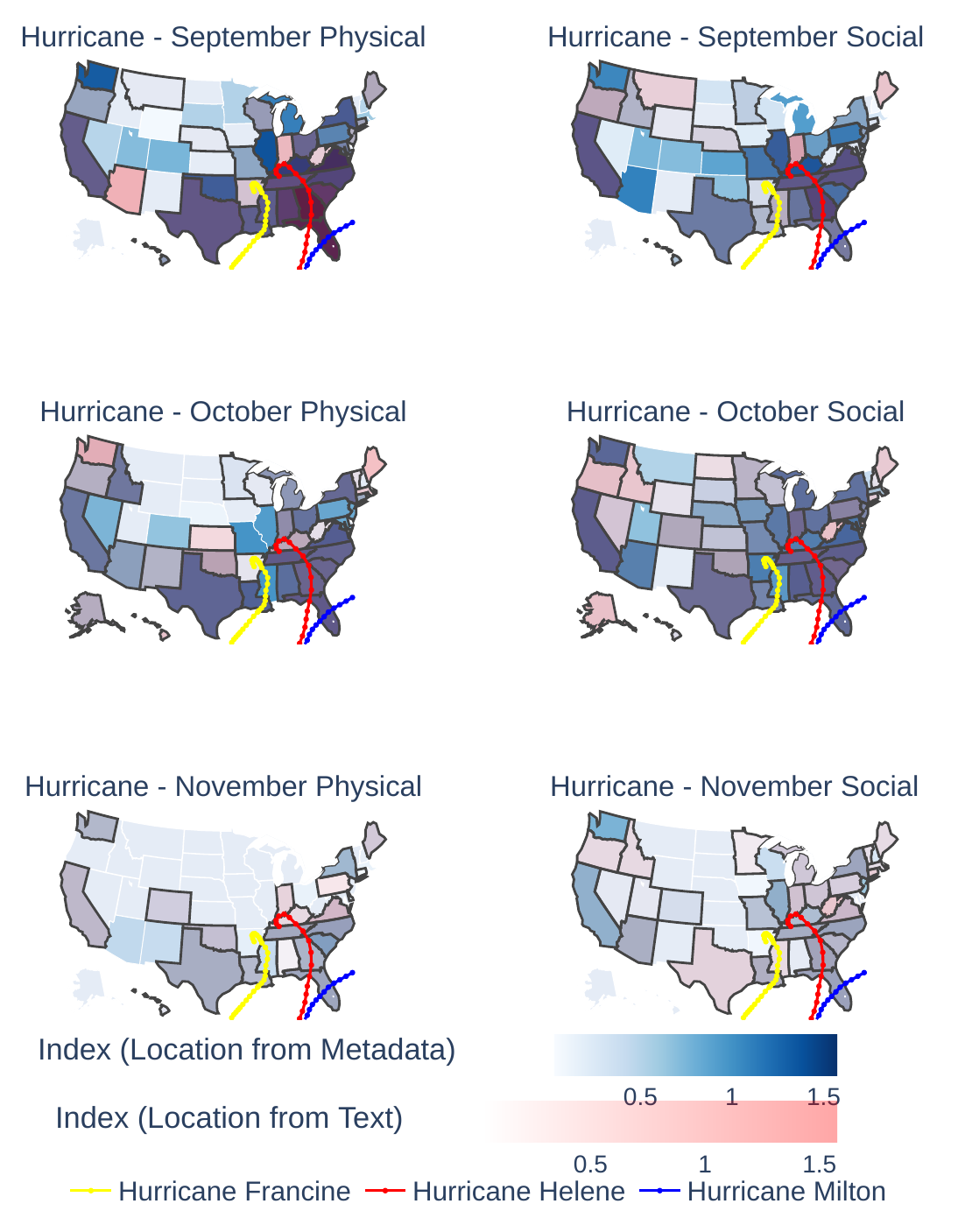} % 图片宽度限制为 minipage 宽度
        \caption{Physi-Social Impact during Hurricanes} % 设置标题
        \label{fig:hurricane_map} % 设置图片引用标签
    \end{minipage}
\end{figure}

\begin{figure}[t]
    \centering
    \begin{minipage}{0.5\textwidth} % 设置整个区域的宽度
        \centering
        \includegraphics[width=0.82\textwidth]{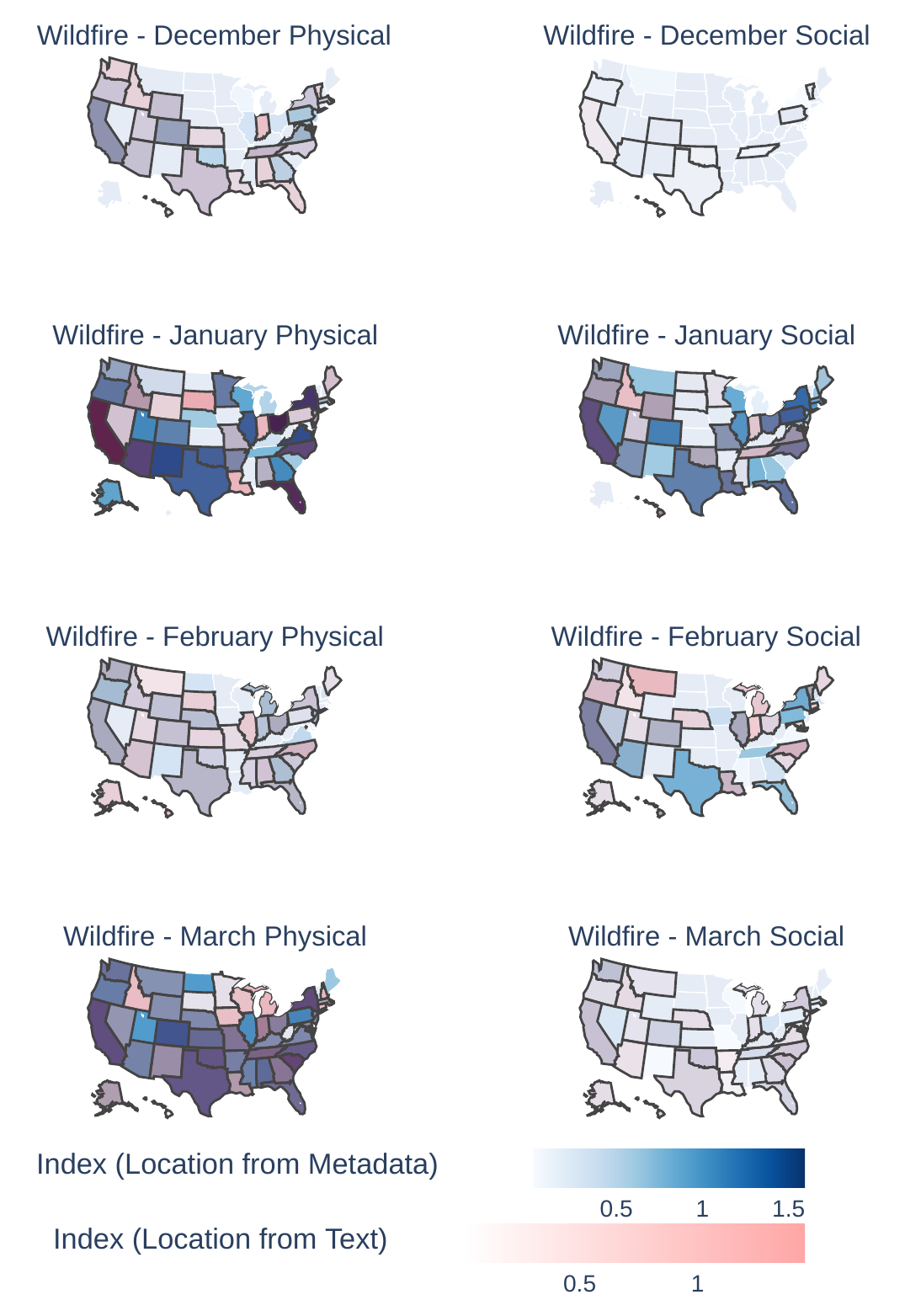} % 图片宽度限制为 minipage 宽度
        \caption{Physi-Social Impact during Wildfires} % 设置标题
        \label{fig:wildfire_map} % 设置图片引用标签
    \end{minipage}
\end{figure}

Figures \ref{fig:hurricane_map} and \ref{fig:wildfire_map} present the monthly average physical and social impact indices for each U.S. state during the hurricane and wildfire periods, respectively. These two approaches correspond to different sources of spatial information. One captures the location where a post was published, while the other identifies the location referenced in the post content. We therefore visually distinguish them as separate layers in the spatial analysis. Specifically, locations inferred from metadata are shown in blue, while locations inferred from textual content are shown in red and overlaid on top of the metadata-based layer for visualization. In addition, we also integrate the posts with geolocation from both methods and the results are available at Appendix Figures \ref{fig:hurricane_map_comb} and \ref{fig:wildfire_map_comb}. Through the analysis, we observe that the physical impact index tends to be significantly higher than the social impact index during the corresponding month in disaster-affected states.
For example, in the hurricane case, states along the storm trajectory, such as Florida, North Carolina, and Georgia, exhibited markedly elevated physical impact indices in September.
Similarly, in the wildfire case, California showed a substantially higher physical impact index in January, reflecting its central role in the wildfire events.
This finding suggests that the spatial distribution of physical impact aligns closely with the actual geographic footprint of the disaster, validating the effectiveness of our framework in capturing region-specific disaster effects.
Moreover, we find that after the disaster, states outside the directly affected regions often exhibit higher social impact indices than physical ones.
This is likely because, although these states are not physically impacted, users in those states still actively engage in discussions around social consequences, such as assistance and recovery efforts.

% \blue{
% We further quantify cross-platform and cross-modality divergence using Jensen–Shannon divergence (Appendix), showing that disaster narratives fragment substantially across platforms and modalities. This analysis spans category distributions across Reddit, TikTok, and YouTube, as well as text, image, and video modalities, motivating the need for unified multi-platform aggregation.
% }

\section{Conclusion}

This paper introduces \textbf{DisImpact}, a two-stage framework for quantifying the physicosocial impacts of disasters using multimodal content from multiple social media platforms. In stage one, an MLLM classifies posts into ten predefined disaster impact categories. In stage two, we compute category-specific indices within each time window using a smoothed proportion of posts and a window-level weight reflecting overall disaster-related activity. Since all indices share a unified scale, they can be compared directly or aggregated into broader physical and social impact measures, enabling both fine-grained category analysis and high-level assessments of overall disaster effects. 
We apply DisImpact to the \textit{2024 Atlantic Hurricanes} and \textit{2025 California wildfires}. Validation against FEMA and NASA FIRMS ground-truth data further demonstrates that the resulting impact indices are timely and reliable. Temporal analysis shows that physical impact indices peak during disaster onset, while social impact indices rise in the aftermath, reflecting shifts in public attention from immediate damage to longer-term social concerns. We also find consistent physical impact patterns across platforms and substantial variation in social impacts, highlighting the value of multi-platform integration.

\section{Limitations and Future Work}

Although DisImpact provides a novel and effective framework for quantifying physi-social impacts of disasters, several limitations remain. 
% At the same time, these limitations highlight natural directions for future research.
\begin{enumerate}
    % \item \textbf{Language and Cultural Coverage}: The current analysis is restricted to posts in English, excluding perspectives from linguistically diverse or marginalized communities. Extending the framework to multilingual data would reduce cultural bias.
    \item \textbf{Single-Label Annotation}: Our framework adopts a single-label annotation scheme that assigns each post to its most dominant impact category following the setting of prior works~\cite{humaid, crisismmd}. However, it loses information about posts that might contain multiple impact types. Future work could extend DisImpact to multi-label annotation and more compositional impact representations, enabling finer-grained modeling for disaster impacts.
    \item \textbf{Data Source}: We also acknowledge that using social media data alone cannot fully capture population-level public sentiment or objective disaster impact. For example, our study finds weak signals of psychological distress. This underrepresentation may reflect both underreporting on public platforms and the long-term nature of mental health impacts. Future work should incorporate complementary data sources (e.g., surveys, NGO reports, or community records) to capture more information.
    % Our study finds weak signals of psychological distress, even though survey-based research documents such effects. This underrepresentation may reflect both underreporting on public platforms and the long-term nature of mental health impacts. Future work should expand the temporal window and incorporate complementary data sources (e.g., surveys, NGO reports, or community records) to capture delayed or less visible consequences.
    \item \textbf{Temporal Scope}: By focusing on posts within a few months before and after disaster events, our framework captures only short-term impacts. Extending the temporal horizon would allow analysis of longer-term recovery processes, including economic resilience, community rebuilding, and mental health outcomes.
    % \item \textbf{Practical Deployment}: Although our current implementation demonstrates research value, translating DisImpact into real-time dashboards or decision support systems could enhance its applied value. Such tools would provide emergency managers and policy makers with early warning signals, support resource allocation, and enable the tracking of recovery trajectories.
    \item  \textbf{Scope of the Unified Index:} Although the unified index enables direct comparison and aggregation across impact categories, it is not intended to replace domain-specific metrics used by practitioners, which often provide higher precision and interpretability within specialized contexts. 
In practice, the index should be interpreted in conjunction with its underlying category-level composition. While experts from different domains may observe the same index value, examining the category breakdown reveals whether observed changes are driven primarily by physical impacts, social impacts, or their interaction. Accordingly, the unified index supports cross-dimensional and cross-platform comparison without collapsing domain-specific meaning.

\end{enumerate}

\bibliography{aaai2026}
% \clearpage
\appendix

% \FloatBarrier   % �� 到这儿，前面所有图必须出现在这里之前

\begin{table*}[t!]
\centering
\footnotesize
\begin{tabular}{c cc|cc}
\toprule
 & \multicolumn{2}{c|}{\textbf{Hurricane}} 
 & \multicolumn{2}{c}{\textbf{Wildfire}}  \\
% \cmidrule(lr){2-3} \cmidrule(lr){4-5}
 & Physical Domain & Social Domain 
 & Physical Domain & Social Domain \\
\midrule
All Relevant Posts 
 & 25,945 (57\%) & 19,281 (43\%)
 &  13,448 (58\%) & 9,903 (42\%)
 \\
 \midrule
Posts with Location from Metadata
& 8,792 (62\%) & 5,339 (38\%)
& 2,671 (64\%) & 1,493 (36\%)
\\
Posts with Location from Textual Content 
& 5,852 (60\%) & 3,836 (40\%)
& 5,507 (56\%) & 4,254 (44\%)
\\
\bottomrule
\end{tabular}
\caption{Distribution of Posts in Physical and Social Domain for Different Settings.}
\label{tab:locationVSall}
\end{table*}

\begin{figure*}[ht]
    \centering
    \begin{minipage}{\textwidth} % 设置整个区域的宽度
        \centering
        \includegraphics[width=0.95\textwidth]{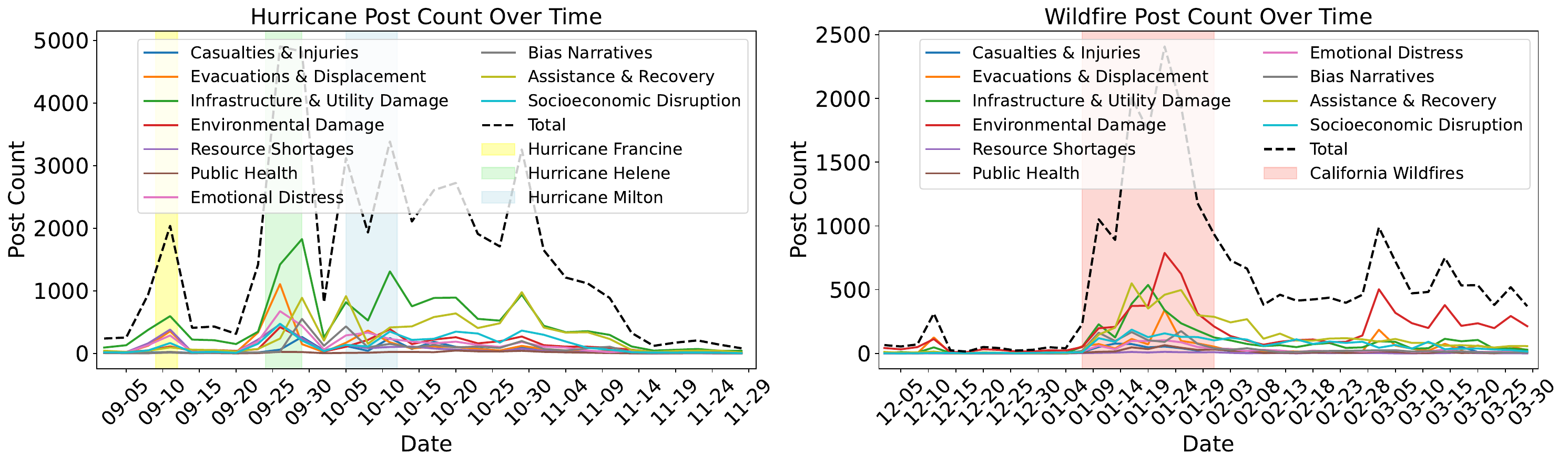} % 图片宽度限制为 minipage 宽度
        \caption{Post Count of each Category Over Time} % 设置标题
        \label{fig:post_count} % 设置图片引用标签
    \end{minipage}
\end{figure*}

\begin{figure}[ht]
    \centering
    \begin{minipage}{0.48\textwidth} % 设置整个区域的宽度
        \centering
        \includegraphics[width=0.95\textwidth]{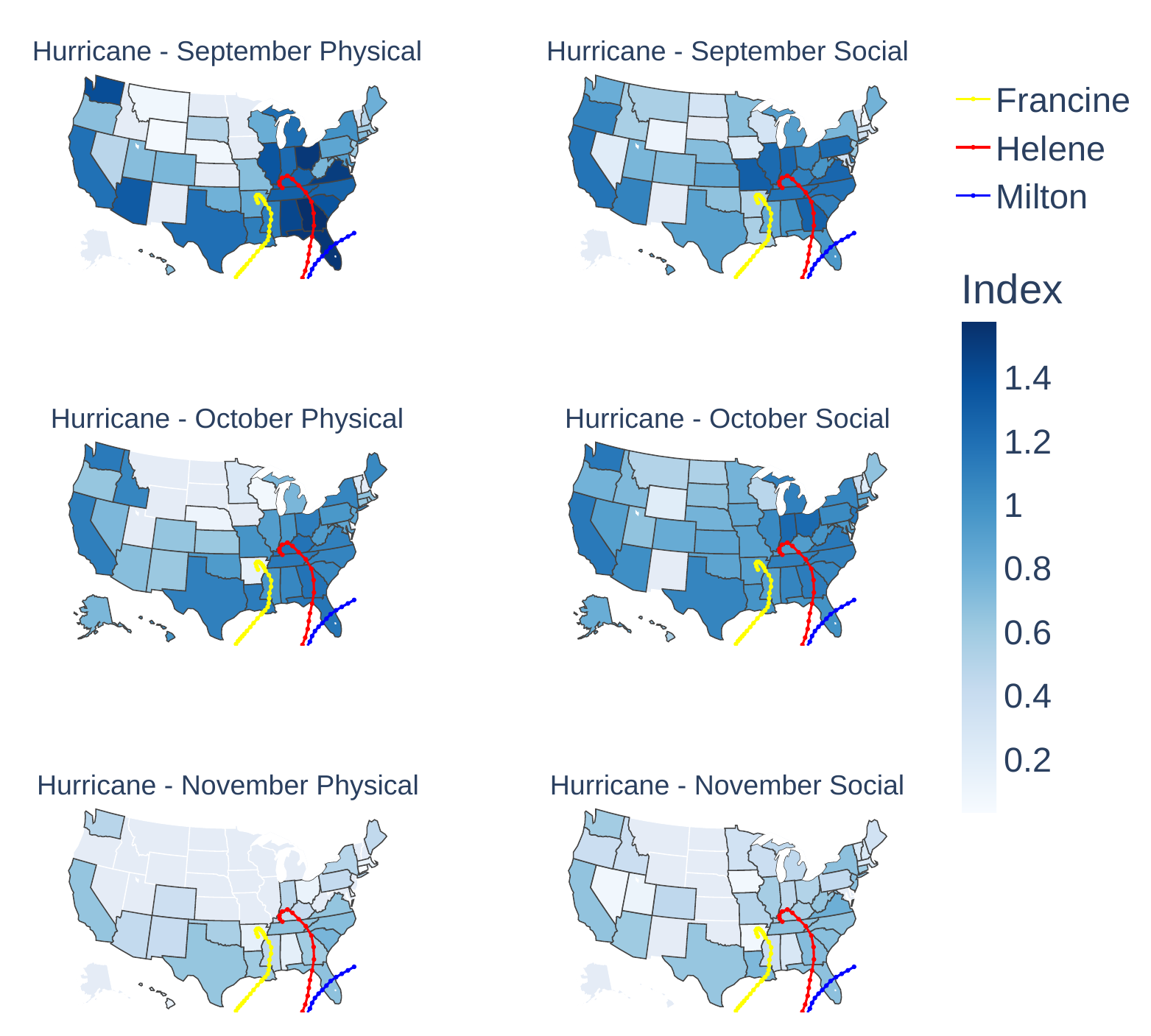} % 图片宽度限制为 minipage 宽度
        \caption{Physi-Social Impact Across U.S. during Hurricanes} % 设置标题
        \label{fig:hurricane_map_comb} % 设置图片引用标签
    \end{minipage}
\end{figure}

\begin{figure}[ht]
    \centering
    \begin{minipage}{0.48\textwidth} % 设置整个区域的宽度
        \centering
        \includegraphics[width=0.95\textwidth]{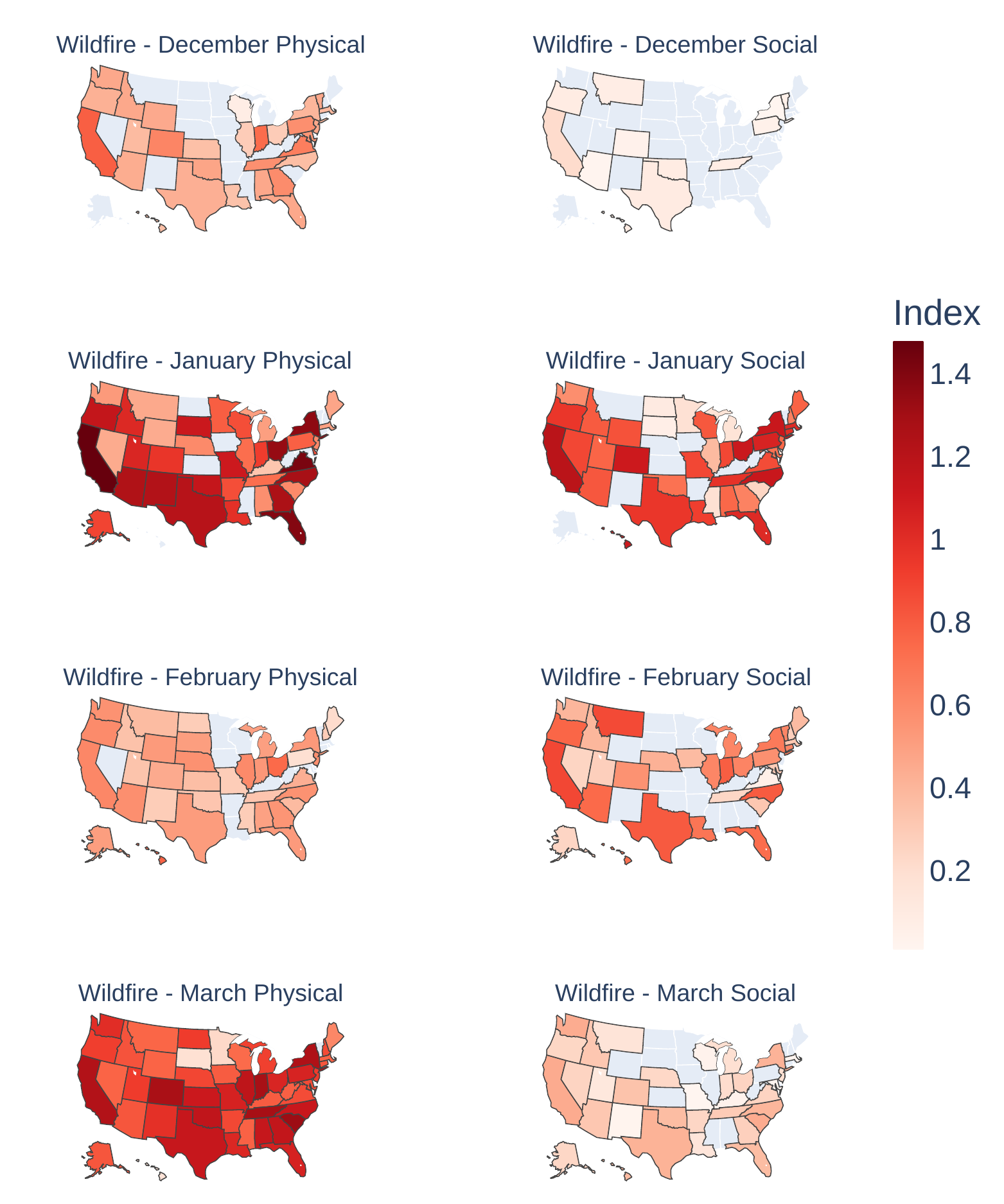} % 图片宽度限制为 minipage 宽度
        \caption{Physi-Social Impact Across U.S. during Wildfires} % 设置标题
        \label{fig:wildfire_map_comb} % 设置图片引用标签
    \end{minipage}
\end{figure}

\begin{figure*}[ht]
\centering
\footnotesize
\begin{tcolorbox}[colback=gray!5, colframe=gray!80!black, title=Prompt]
Read the post, considering text, image, and video together. Determine whether the post is related to an actual hurricane disaster in North America (especially Hurricane Helene and Hurricane Milton) and give your reason. \\ \\
Accepted examples:\\
- Hurricane disaster, even if it's not Helene or Milton\\ \\
Some typical counter examples:\\
- Miami Hurricanes football team, Carolina Hurricanes hockey team, or other sports names\\
- WWE wrestler called ‘Hurricane’ or other people called 'Hurricane'\\
- Advertisement\\
- Use hurricane as a metaphor to describe other things\\
- The post contains hurricane-related hashtag, but the actual content is not related to the hurricane disaster\\
- Cyclone or Typhoon that does not happen in North America\\ \\
Respond with:\\
\{``Judgment'': True or False
\}

\end{tcolorbox}
\caption{Prompt for Hurricane Data Cleaning.}
\label{fig:prompt-clean-hurricane}
\end{figure*}
% \FloatBarrier   % �� 到这儿，前面所有图必须出现在这里之前

\begin{figure*}[ht]
\centering
\footnotesize
\begin{tcolorbox}[colback=gray!5, colframe=gray!80!black, title=Prompt]
Read the post, considering text and video together. Determine whether the post is related to a wildfire disaster in North America (especially LA wildfires) and give your reason. \\ \\  
Accepted examples:  \\
- Wildfire disaster in North America, even if it is not in California  \\ \\
Some typical counter examples:  \\
- Sports teams, movies, songs, or products with “Wildfire” in the name  \\
- Use of “wildfire” as a metaphor (e.g., “the rumor spread like wildfire”)  \\
- Wildfire events clearly located outside North America (e.g., Australia, Greece)  \\
- Other disasters such as hurricanes, floods, earthquakes, or tornadoes  \\
- Posts with wildfire-related hashtags, but actual content not about the wildfire disaster\\
- Advertisements or promotions unrelated to wildfire disasters\\

Respond with:\\
\{``Judgment'': True or False
\}

\end{tcolorbox}
\caption{Prompt for Wildfire Data Cleaning.}
\label{fig:prompt-clean-wildfire}
\end{figure*}
% \FloatBarrier   % �� 到这儿，前面所有图必须出现在这里之前

\begin{figure*}[ht]
\centering
\footnotesize
\begin{tcolorbox}[colback=gray!5, colframe=gray!80!black, title=Prompt]
You are a disaster-focused social media classification assistant. Your task is to analyze social media posts and determine which one of the predefined impact categories best describes the post, based on its text, image, and video content. If none of the categories apply, return 11 (Other / Not relevant).\\ \\  
Please select the single most appropriate category from the list below.\\ \\
- Casualties \& Injuries (1): Posts describing people or animals who are killed, seriously injured, missing, or experiencing immediate medical emergencies as a result of the disaster.\\
- Evacuations \& Displacement (2): Posts describing people being forced to evacuate, or relocations to shelters caused by unsafe conditions during a disaster.\\
- Infrastructure \& Utility Damage (3): Posts reporting physical damage to infrastructure or utility systems, such as roads, buildings, bridges, or disruptions to essential services such as electricity, water, or communication networks.\\
- Environmental Damage (4): Posts describing harm to the natural environment or resources caused by the disaster, such as damage to ecosystems, agriculture, or coastlines, as well as contamination of water, soil, or air.\\
- Resource Shortages (5): Posts requesting essential survival resources during a disaster, such as clean water, food, shelter, clothing, or other urgent supplies.\\
- Public Health (6): Posts describing public health consequences following a disaster, such as outbreaks of infectious diseases or mold-related illnesses, disruption of chronic illness care, and shortages of medical supplies or hospital services.\\
- Emotional and Psychological Distress (7): Posts describing psychological distress or emotional suffering experienced by oneself or others as a result of a disaster, such as trauma, anxiety, depression, grief, or cumulative emotional strain.\\
- Bias Narratives (8): Posts revealing social bias, discrimination, or unequal impacts of the disaster across different groups, such as political blame, hate speech, or highlighting the disproportionate suffering of marginalized or vulnerable populations.\\
- Assistance \& Recovery (9): Posts describing efforts to support recovery and rebuilding after a disaster, such as formal aid from governments or NGOs, informal community-based assistance, long-term infrastructure repair, and expressions of resilience or adaptation by affected communities.\\
- Socioeconomic Disruption (10): Posts describing economic or social disruptions caused by the disaster, such as financial losses, business interruptions, housing insecurity, educational setbacks, job loss, or the decline of local industries such as tourism and retail.\\
- Other / Not Relevant (11): None of the above categories apply to this post.
Return only the number of the most appropriate category (1-11). Do not include explanations unless asked.\\

Respond with:\\
\{``Judgment'': Your Judgment (1-11 integer)
\}

\end{tcolorbox}
\caption{Prompt for Physi-Social Classification.}
\label{fig:prompt-classification}
\end{figure*}

\end{document}